\documentclass[prl,twocolumn,showpacs,preprintnumbers]{revtex4}
\usepackage{amsmath,amssymb,latexsym}

\usepackage{graphicx}

\begin{document}

\title
{Current issues in finite-$T$ density-functional theory and Warm-Correlated Matter$^\S$.

}
$^\S${\it Presented at the DFT16 meeting in Debrecen, 2015, held on the
50$^{th}$ anniversary of Kohn-Sham Theory.}

\author
{
 M.W.C. Dharma-wardana}
\affiliation{
National Research Council of Canada, Ottawa, Canada, K1A 0R6.}
\email[Email address:\ ]{chandre.dharma-wardana@nrc-cnrc.gc.ca}
\affiliation{
D\'{e}partement de Physique, Universit\'{e} de Montr\'{e}al.
}

%
\date{\today}
%

\begin{abstract} 
Finite-temperature DFT has become of topical interest, partly due to the
increasing ability to create  novel states
of warm-correlated matter (WCM). Warm-dense matter (WDM), ultra-fast
matter (UFM), and high-energy density matter (HEDM) may all be regard as
subclasses of WCM. Strong electron-electron, ion-ion and electron-ion
correlation effects and partial degeneracies are found in these systems where
the electron  temperature $T_e$ is comparable to the electron Fermi
energy $E_F$. Thus many electrons are in continuum states which are partially
occupied. The ion subsystem may be solid, liquid or plasma, with many states of
ionization with ionic charge $Z_j$. Quasi-equilibria with the ion temperature
$T_i\ne T_e$ are common. The ion subsystem in WCM can no longer be treated as a
passive ``external potential",  as is customary in $T=0$ density functional
theory (DFT) dominated by solid-state theory or quantum chemistry.
Many basic questions arise in trying to implement DFT for WCM.
Hohenberg-Kohn-Mermin theory can be adapted for treating these systems if
suitable finite-$T$ exchange-correlation (XC)  functionals can be constructed.
They  are functionals of both the one-body electron density $n_e$ and the
one-body ion densities $\rho_j$.  Here $j$ counts many species of nuclei or
charge states. A method of approximately but accurately mapping the
 quantum electrons to a classical Coulomb gas enables one to treat electron-ion
systems entirely classically at any temperature and arbitrary spin polarization,
 using exchange-correlation effects
 calculated {\it in situ}, directly from the pair-distribution functions.
 This  eliminates the need for any XC-functionals. This classical map has
 been used to calculate the equation of state of WDM systems, and 
construct a finite-$T$ XC functional that is found to be
in close agreement with recent quantum path-integral simulation data.
In this review current developments and concerns in finite-$T$ DFT, especially
in the context of non-relativistic  warm-dense matter and ultra-fast matter will
be presented.
\end{abstract}
%
\maketitle

\section{Introduction}
\label{intro} 
Although there are  no systems at zero temperature available to us, it is the
quantum mechanics of the simpler $T$=0 systems that has engaged  the attention
of theorists. Thermal ensembles usually require the study of extended systems
attached to a ``heat bath'', and  within some statistical ensemble. Even
perturbation-theory   approaches to model systems like the electron gas at
finite-$T$ were full of surprises~\cite{Luttinger60}.

 Condensed matter physics and chemistry could get by with $T=0$ quantum
mechanics as the input to some sort of thermal theory which is not integrated
into the many-body problem. Much of plasma physics and astrophysics could
manage with simple extensions of hydrogenic models, Thomas-Fermi theory,
extended-Debye theory, and classical  `one-component-plasma' models as long as 
the accuracy of observations, experiments and theoretical models did  not
demand anything more from quantum mechanics. On the other hand, at the level of
foundations of quantum mechanics, the whole issue of  a quantized thermo-field
dynamics  has been an open problem~\cite{Umezawa}. Similarly, the theory of `mixed'
systems with classical and quantum components is also a topic of
discussion~\cite{mixedQ}. It is in this context that we need to look at
the advent of density-functional theory (DFT) as a great step forward in the
quantum many-body problem. The Hohenberg-Kohn theorem published in 1964 was
soon followed by its finite-$T$ generalization by Mermin, providing a `thermal'
density-functional theory (th-DFT) in 1965~\cite{DFT-source, Mermin}, which also saw
the advent of Kohn-Sham theory. Hence in 2015 we are celebrating the fiftieth anniversary
of both Kohn-Sham theory, and Mermin's extension of Hohenberg-Kohn theory
 to finite-$T$.

 While DFT provided chemistry and condensed-matter physics an escape from the
intractable `$n$-electron' problem,  in addition to its 
computational implications, DFT has  deep epistemological implications in regard
to the foundational ideas of physics.  DFT claims that the many-body wavefunction
can be dispensed with, and that the physics of a given system
can be discussed as a functional of the one-body density. Thus even entanglement can be
discussed in terms of density functionals~\cite{apvmm,cdwPisa2013}.
 However, it is the computational power of DFT that has been universally
 exploited in many  fields of physics.

The interest in thermonuclear fusion via laser compression and related
techniques, and the advent of ultra-fast lasers have created novel states of
matter where the electron temperature $T_e$ is usually of the order of the
Fermi energy $E_F$, under conditions where they are identified as
warm dense matter (WDM)~\cite{graz-Desj-Red2014}.   When WDM is  created
using a fast laser within femto-second time-scales, the photons couple strongly
to the electrons which are  heated very rapidly to many thousands of degrees,
while the ions remain essentially at the initial `ambient'
temperature~\cite{Milsch,Ng2011}.  In addition to highly non-equilibrium systems,
this often leads to two-temperature systems with the ion
temperature $T_i\ne T_e$, with $T_e\gg T_i$. Alternatively, if shock waves are
used to generated a WDM we may have $T_i>T_e$. Such ultra-fast matter (UFM)
 systems can be studied
using a fs-probe laser within timescales $t$ such that $t\ll \tau_{ei}$, where
$\tau_{ei}$ is the electron-ion temperature relaxation time~\cite{elr} of the
UFM system.   These WCM systems are of interest in astrophysics and planetary
science~\cite{Militzer}, inertial fusion~\cite{graz-Desj-Red2014}, materials
  ablation and
machining~\cite{ablation}, in the hot-carrier physics of field-effect
transistors  and other nano-devices~\cite{JagdeepShah}.

Early attempts to apply thermal-DFT (also called finite-$T$ DFT, th-DFT)
 to WDM-like
systems were undertaken by the present author and Fran\c{c}ois Perrot in the
early 1980s as reviewed in Ref.~\cite{cdw-CPP}. This involved a
 reformulation of the neutral-pseudoatom (NPA) model that
had been formulated by Dagens~\cite{Dagens} for zero-$T$ problems, 
as it has the versatility
to treat solids, liquids and plasmas. 

Originally it was J. M. Ziman~\cite{Ziman}
 (and possibly others) who  had
proposed the  NPA model as an intuitive physical idea in the context of
solid-state physics. The electronic structure of matter is regarded  as a
superposition of charge densities $n_j(\vec{r}-\vec{R}_j)$ centered on each
nuclear center at $\vec{R_j}$. In other words, if the total charge density in
momentum space was $n_T\vec{(k})$, then this is considered as being made up of
the individual charge distributions $n_j(\vec{k})$ put together using the ionic
structure factor $S(\vec{k})$. This was more explicitly implemented in
muffin-tin models of solids, or `atoms-in molecules' models of chemical bonds
that were actively pursued in the 1960s, with the increasing availability of
fast computers. The NPA model was formulated rigorously within $T=0$ DFT by
Dagens who showed that it was capable of the same level of accuracy, at least
for `simple metals'  as the LMTO, APW or Korringa-Kohn-Rostoker codes  that
were becoming available in the 1970s~\cite{Dagens}. Wigner's $T=0$ exchange-
correlation (XC) `functional' in the local-density approximation (LDA) was used
by Dagens.   

In the finite-$T$ NPA that we have used
as our ``work-horse", we solve the Kohn-Sham Mermin equation for a single nucleus
placed at the center of a large `correlation sphere' of radius $R_c$ which is of the order
of 10$r_{ws}$, where $r_{ws}$ is the Wigner-Seitz radius per ion. 
Here $r_{ws}=\{3/(4\pi \bar{\rho})\}^{1/3}$ where $\bar{\rho}$ is the
ion density given as the number of ions per unit atomic volume. For WDM aluminum
at normal compression $r_{ws}\simeq 3$ a.u. All types of particle
correlations induced by the nucleus at the center of the `correlation sphere' 
would have died down to bulk-values when $r\to R_c$. The ion distribution
$\rho(r)=\bar{\rho}g_{ii}(r)$ is approximated as a spherical cavity of radius
$r_{ws}$ surrounding the nucleus, and then becoming a uniform
 positive background~\cite{PerrotBe, pdw-Al}. This is simpler to implement than
 the full method implemented  in Ref.~\cite{dwp82}. The latter involved
 a self-consistent
iteration of the ion density $\rho(r)$ and the electron density $n(r)$ 
obtained from the Kohn-Sham procedure coupled to a classical integral
equation or even molecular dynamics;  the simpler NPA procedure is sufficient
 in most cases.

There have also been several practical formulations of NPA-like models in more
recent times. Some of these~\cite{Wilson}  are extensions of the INFERNO
cell-model of Lieberman~\cite{Lieberman}, while others~\cite{StarrSau}
  use a mixture of NPA
ideas as well as elements of Chihara's "quantum-HNC" models~\cite{Chihara94}.
We have discussed  Chihara's model to some extent in
Ref.~\cite{Sanibel11}. In true DFT models the electrons are mapped to a  
non-interacting Kohn-Sham electron gas having the same interacting density but
at the non-interacting chemical potential. This feature is absent in
INFERNO-like  cell-models where the chemical potential is determined via
an integration within the ion-sphere or by some such consideration.
Thus different physical results may arise (e.g., for the conductivity)
 depending on how the chemical potential is fixed.  Chihara's models use
 an ion susbsytem and  an electron subsystem coupled via a `quantal
 Ornstein-Zernike' equation.  However, if a  one-component electron-gas
 calculation
were attempted via the `quantal HNC', the known $g_{ee}(r)$ are not recovered.
In the two component case, as far as we can ascertain, the $S(k\to 0)$
 limit is not correctly related to the compressibility.

Thus the Kohn-Sham NPA calculation provides the free-electron charge
 density pile-up $n_f(r)$
around the nucleus. This is sufficient to calculate an electron-ion pseudopotential
$U_{ei}$, and hence an ion-ion pair potential $V_{ii}(r)$    as discussed in, say,
Ref.~\cite{pdw-Al}. Once the pair-potential is available, the Hyper-Netted Chain
equation (and its modified form incorporating a bridge function) can be used to
calculate an accurate $g_{ii}(r)$ if desired, rather than via the
direct iterative procedure used in Ref.~\cite{dwp82}. This finite-$T$ NPA approach
 is capable of accurate prediction of phonons (i.e., milli-volt energies)
in WDM systems, as shown explicitly by Harbour {\it et al.}~\cite{LouisCPP} using
 comparisons with results  
 reported by Recoules {\it et al.}~\cite{Recoules}
who used the Vienna Ab-initio  Simulation
Package (VASP). 

.

Since the XC-functional of DFT is directly connected with the pair-distribution
function (PDF), or equivalently with the two-particle density
matrix~\cite{Gilbert75}, we sought to formulate the many-body problem of
ion-electron systems directly in terms of the pair distribution functions
$g_{\alpha,\beta}$ of the system, where $\alpha$  and $\beta$ count over types
of particles (ions and electrons, with two types of electrons with spin up, or
down)~\cite{dwp82, PerrotBe, pdw-Al}.  The ionic species may be  regard as
classical particles without spin as their thermal de Broglie length is in the
femto-meter regime at WDM temperatures. This approach led to the formulation of
the Classical-map Hyper-Netted-chain (CHNC) method that will be briefly
described in sec.~\ref{chnc.sub}. 

The attempt to use thermal DFT for actual calculations naturally required an
effort towards the development of finite-$T$
XC-functionals~\cite{DWT,RajGupt,PDW84,Callaway, Dandrea, Ichimaru, pdw2000}.
Meanwhile, large-scale codes implementing $T=0$ DFT (e.g., CASTEP~\cite{castep},
 VASP~\cite{vasp}, ABINIT~\cite{abinit}, ADF~\cite{adf}, 
Gaussian~\cite{gaussian} etc.) became available, where well-tested  $T=0$
XC-functionals (e.g., the PBE functional~\cite{PBE})  as well as $T=0$
DFT-based pseudopotentials are implemented.  Currently, these codes also
included versions where the single-particle states could be chosen as a Fermi
distribution~\cite{Gonze} at a given temperature, while they do not include the
finite-$T$ XC functionals that are needed for a proper implementation of
thermal DFT. 
  These codes are meant to be used at $T=0$ or small
$T$ since finite-$T$ calculations require  a very rapid increase in the basis
sets needed for such calculations. 
It should also be mentioned that Karasiev {\it et al.}l~\cite{KaraQE}
have recently implemented finite-$T$ XC within the `Quantum Espresso' code,
as well as given an "orbital-free" implementation, although, as far as we can see,
the non-locality problem in the kinetic-energy functional has not been resolved.

However, the availability of DFT-electronic structure  codes
have opened up the possibility of using them  even in the  WDM  regime. We give
several  references to such work  which
contain additional citations to other calculations
\cite{silv1,Recoules,Pleg15,Sper15,Vinko15,CDW-lcls}.  This renewed interest has
re-kindled an interest in the theory of thermal DFT  in the context of current
concerns~\cite{ppgb13}. In the following we discuss  some of the typical issues
that arise in  applying thermal-DFT to current problems, as these may range
from basic issues to the simple question of ``if one can get away with" just
using the $T=0$ XC functional.

The use of a functional, augmented with gradient approximations etc, is
satisfactory as long as the `external potential' can be considered fixed, as is
the usual case in quantum chemistry and solid-state physics. In situations
where the external potential arises from a dynamic ion distribution $\rho(r)$,
since $\rho(r)$ as well as the electron distribution $n(r)$  depend
self-consistently on each other, it is clear that the XC-contribution is a
functional of both $\rho$ and $n$, i.e., the XC-functional is of the form 
$F[n(r),\rho(r)]$.  Under such circumstances, a direct {\it in  situ}
calculation of the electron $g(r)$ in the presence of  the ion distribution has
to be carried out, and an `on-the-fly' coupling constant integration is needed 
for each self-consistent loop determining $n(r)$ and $\rho(r)$. We  presented
examples of such calculations for a system of electrons and protons at finite
temperatures, in Refs.~\cite{hug,ppots12}, using the  classical-map
Hyper-netted Chain technique (CHNC) that enables an easy {\it in situ} 
calculation of the $g_{ee}(r), g_{ei}(r)$ and $g_{ii}(r)$. This approach is
at once non-local and hence avoids the need for gradient approximations.
Furthermore, the ion-ion correlations are highly non-local and the LDA
or its extensions are totally inadequate since they are described by the
HNC approximation.

\begin{figure}[t]
\vspace{0.5cm}
\centering
\includegraphics[width=8cm]{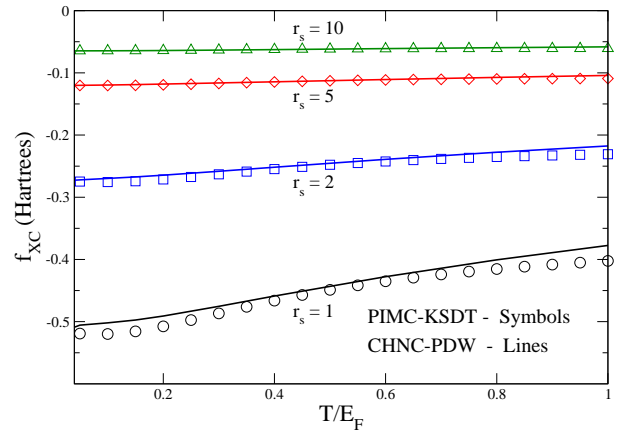}
\caption{(Color online) 
Finite-$T$ exchange and correlation free energy $f_{xc}(r_s,T)$ per electron
(Hartrees) versus the reduced temperature $T/E_F$ in units of the Fermi energy.
The symbols, labeled PIMC-KSDT are the fit given by Karasiev {\it et al.},
Ref.~\cite{KSDT} to the path-integral Monte Carlo (PIMC) data of Brown {\it et al.},
Ref.~\cite{Brown13}. The continuous lines, labeled CHNC-PDW are from the
classical-map HNC procedure of Perrot and Dharma-wardana, Ref.~\cite{pdw2000}.
The temperature range $0<T/E_F \le 1$ is the region of interest for WDM studies. 
}
\label{fxc-fig}
\end{figure}

\section{Exchange-correlation at finite-$T$}
\label{xc-sec}
It may be useful to present this section as an `FAQ' (frequently asked
questions) rather than a formal discussion on thermal-XC functionals.
\subsection{Do we have reliable thermal-XC functionals?}
\label{xc-func.sub}
 The finite-$T$ XC-functional in the random-phase approximation
(RPA)~\cite{RajGupt,PDW84,Callaway} has been available since 1982, while
formulations and parametrizations that go beyond RPA have been available since
the late 1980s~\cite{Dandrea,Ichimaru,pdw2000}. Finite-$T$  XC-data from
quantum simulations for the uniform finite-$T$  electron fluid were provided 
in 2013 by Brown {\it et al.}~\cite{Brown13}, while an analytical fit to their data is
found in Karasiev {\it et al.}~\cite{KSDT}. The XC-parametrization of Perrot and
Dharma-wardana given in 2000, Ref.~\cite{pdw2000}, was based on a coupling-constant
evaluation of the finite-$T$ electron-fluid PDF calculated via the
Classical-map Hyper-Netted-Chain (CHNC)~\cite{prl1} method. It  closely agrees
with the recent quantum-simulation results (Fig~\ref{fxc-fig}). 
Finite-$T$ CHNC-based
results are available for the 2D- and 3D- electron gas,  as well as other
electron-layer systems~\cite{prl3,2v2d, thick2d}. They are in good agreement
with path-integral and other Monte Carlo (PIMC) calculations where available.
 
We consider
the data for the 3D system that have been conveniently parametrized by Karasiev
{\it et al.},  (labeled KSDT   in Fig.~\ref{fxc-fig}). The CHNC $f_{xc}(T)$ at high
temperatures (beyond what is displayed in the figure) show somewhat less
correlation than given by PIMC, but correctly approaches the Debye-H\"{u}ckel
limit at high temperatures. In the high-density regime ($r_s<1$), the
RPA-functionals become increasingly accurate as $r_s\to 0$. The
small-$r_s$ regime has also been recently treated by Schoof {\it et al.}~\cite{Schoof}.
It should be stated that when the CHNC mapping was constructed,
 Fran\c{o}is Perrot and the
present author did not attempt to map  the  $r_s < 1 $ regime in detail as
it is fairly well treated by RPA methods.
 Recent simulations by Malone {\it et al.}~\cite{Malone} find some differences between
 their work, and that of Brown at al for $r_s$ in the neighbourhood of unity. 
Similarly, the CHNC
data show  differences for the $r_s=1$ curve,  as shown in Fig~\ref{fxc-fig}.
However it is too early to re-examine the small $r_s$ regime and
review the  data of Ref.~\cite{Malone} which are given as the internal energy and
not converted to a free energy.  

However, it is clear that there is no shortage of reliable finite-$T$ XC-functionals
 for those who wish to use them.
\subsection{Can we ignore thermal corrections and use the $T=0$ implementations?}
\label{ignore.sub}
While finite-$T$ XC functionals can be easily incorporated into the NPA model
or  average-atom cell models etc.~\cite{Lieberman}, this is much more difficult
in the context of large DFT codes like VASP or ABINIT. Hence the already
installed $T=0$ XC-functionals have been used as a part of the `package' for a
significant number of  calculations for WDM materials, ranging from equation of
state (EOS),  X-ray Thomson scattering, conductivity etc. Hence the question
has been raised as to whether the thermal corrections to the $T=0$
XC-functional may be conveniently  disregarded. 

The push for accurate XC-functionals in quantum chemistry came from the need
for `chemical accuracy' in predicting molecular interactions in the
milli-Rydberg range. The current level of accuracy in WDM experiments is
nowhere near that. Furthermore, many properties (e.g., the EOS and the total
energy) are insensitive to details since total energies are usually very large
compared to XC-energies, even at $T=0$, unless one is dealing with unusually
contrived few-particle
systems. However, one can give a number of counter examples which are designed
to show that there are many situations  where the thermal modification of the
$T=0$ XC-energy and XC-potential are important.

 As a model system we may consider the uniform electron fluid with a density of
$n$ electrons per atomic volume, and thus having an electron-sphere radius
$r_s=\{3/(4\pi n)\}^{1/3}$. Since the Fermi momentum $k_F=1/(\alpha r_s)$,
where $\alpha=(4/9\pi)^{1/3}$, the kinetic energy at $T=0$ scales as $1/r_s^2$,
while the Coulomb energy scales as $1/r_s$. Hence the ratio of the
Coulomb-interaction  energy to the kinetic energy scales as $r_s$.  Thus, the
electron-sphere radius $r_s$ is also the `coupling constant' that indicates the
deviation of the system from the non-interacting independent particle model.
The RPA is valid when $r_s < 1$ for $T=0$ systems, for  Coulomb
fluids.  On the other hand, at very high temperatures, the kinetic energy
becomes $T$ (or $k_BT$ where $k_B=1$ in our units), while the Coulomb energy is
$Z^2/r_s$, where $Z=e=-1$ for the electron fluid. Hence the ratio of the
Coulomb energy to the kinetic energy, viz., $\Gamma=Z^2/(r_sT)$ for Classical
Coulomb systems. Here the role of  $r_s$ is reversed to that at $T=0$,
 and the system behaves as
an ``ideal gas" for large $r_s$ in systems where $T\gg E_F$. The equivalent of
the RPA-theory  in the high-$T$ limit is the Debye-H\"{u}ckel theory which is
valid for $\Gamma<1$.   A generalized coupling constant that `switches over'
correctly from its $T=0$ behavior to the classical-fluid behaviour at high
$T$ can be given as:
\begin{eqnarray}
\Gamma(r_s,T)&=&P.E/K.E=Z^2/(r_s T_{kin})  \\
\Gamma(r_s,T\to 0)&=&r_s,\;\; \Gamma(r_s,T\to\infty)=Z^2/(r_sT)
\end{eqnarray}
The equivalent kinetic temperature $T_{kin}$ referred to in the above
equation can be constructed from the mean kinetic energy as in Eq.~(A2) given
in the appendix to Ref.~\cite{pdw2000}. However, the main point here is that
there are {\it two non-interacting limits} for studying Coulomb fluids. We can
start from the $T=0$ non-interacting limit and carry out
perturbation theory, or coupling-constant integrations to include the effect of
the Coulomb interaction $\lambda Z^2/r$, with $\lambda$ moving from 0 to unity
(e.g., see Eq. 71 of Ref.~\cite{ppgb13} for a  discussion and
references). Alternatively, we can start from the $T \to \infty$
non-interacting limit. This high temperature limit is the `classical limit' where
the system is a non-interacting Boltzmann gas.
One can do perturbation theory as well as coupling constant integrations over
$\Gamma'$ going from 0 to its required value $\Gamma$. The latter approach is
well known in the theory of classical fluids. Such results provide standard
`benchmarks' in the context of the classical one-component
plasma~\cite{BausHan80,Ichimaru}, just as the electron gas does for the
 quantum many-electron
problem. However, there is no clear way of evolving from a classical Boltzmann gas
at $\Gamma=0$ into a quantum fluid by increasing the Coulomb coupling
 to its full value, as the anti-symmetry of the underlying wavefunction
needs to be included.
This problem does not arise if we start from a non-interacting Fermi
gas at $T=0$. How this problem is solved within a classical scheme is
discussed below, in the context of the CHNC method. The `temperature
 connection formula' referred to recently by Burke {\it et al.}~\cite{BurkeSPB}
 in a thermal-DFT context may be closely related to this discussion.

Although XC effects are important, it is a small fraction of the total
energy. They become negligible as $T$  becomes very large, when the total
energy itself becomes very large. Thus  it is easy to understand that
finite-$T$ XC effects are most important, for any given
$r_s$,  in the WDM range where $0\le T/E_F \le 1$, with $E_F=0.5/(\alpha
r_s)^2$.   Furthermore, in any electron-ion system containing {\it even one}
bound state, the electron density $n(r)$ becomes large as one approaches the
atomic core, and hence there are spatial regions $r$ where $T/E_F(r)\le 1$,
when finite-$T$ XC comes into play. Since the
`free-electron' states are orthogonal to the core states, the free-electron
density pile-up  $n_f(r)$ near a nucleus immersed in a hot-electron fluid is
also equally affected, directly and via the core. Furthermore, $n_f(r)$ is a
property that directly enters into the calculation of the  X-Ray Thomson
scattering signal as well as the electron-ion pseudopotential $U_{ei}(r)$.
 Hence the effect of finite-$T$ XC, and the need to include
thermal-XC functionals in such calculations  can be experimentally ascertained.

\begin{figure}[t]
\vspace{0.5cm}
\centering
\includegraphics[width=8cm]{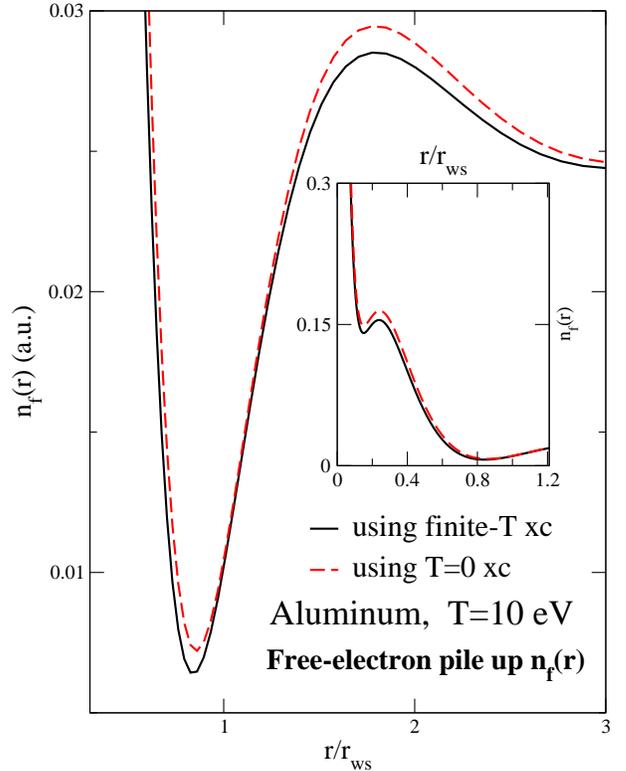}
\caption{(Color online) The NPA free-electron  density $n_f(r)$
using PDW finite-$T$ XC and  with the $T=0$ XC. Inset: $n_f(r)$
 inside the Wigner-Seitz sphere, with
$r_{ws} \simeq$ 3.0 Bohr.}
\label{Thomp-fig}
\end{figure}

In Fig.~\ref{Thomp-fig} we present  the  $n_f(r)$
near an Aluminum
nucleus in an electron fluid of density $1.81\times 10^{23}$ electrons/cm$^3$, i.e.,
at $r_s=2.07$ and  at  $T= 10$ eV, calculated using the neutral-pseudo-atom method.
 This temperature corresponds to $T/E_F\simeq
0.84$.  Calculations using VASP code for an actual experiment covering
 this regime has also been given  by Plageman {\it et al.}~\cite{Pleg15}. 
Although the difference in charge densities that arises from the difference
between the $T=0$ XC and the finite-$T$ XC shown in Fig.~\ref{Thomp-fig} may
seem small, such charge-density differences translate into significant energy
differences as well as into significant  X-ray scattering features. 

Although
Kohn-Sham energies are not to be interpreted as the one-particle excitation
energies of the system, they can be regarded as the one-particle  energies of
the non-interacting electron fluid (at the interacting density) that appears in
Kohn-Sham theory. These eigen-energies are also sensitive to whether we use the
$T=0$ XC-functional, or even to different finite-$T$  functionals. For
instance, in Sec. 6 of Ref.~\cite{pdw2000} we give the Kohn-Sham energy
spectrum of warm-dense Aluminum at 15 eV calculated using the PDW-finite-$T$
XC-functional~\cite{pdw2000}, as well as the finite-$T$ Iyatomi-Ichimaru (YI) 
functional. In summary, the KS-bound states obtained  by the two methods (with
YI given second) are: at energies (in Rydbergs) of 2115.044 and 2110.199 for
the 1s level, 27.86214 and 27.53968 for the 2s level. The outermost level, the
2p-state, has an energy  of 25.05646 and 24.81116 from PWD and YI,
respectively. Similar proportionate changes are seen in the phase shifts of the
continuum states. Thus it is  clear that the XC-potentials  should have a
significant impact, especially in determining the regimes of plasma phase
transitions~\cite{pdw-Al,Norman68}, finite-$T$ magnetic transitions, as well as in the
theory of ionization processes~\cite{Vinko15} and transport properties. 

Another example of the need for finite-$T$ XC functionals is given by Sjostrom
and Daligault~\cite{SjostromDalig14} in their discussion of gradient-corrected
thermal functionals. They conclude that ``finite- temperature functionals show
improvement over zero-temperature functionals, as compared to path-integral
Monte Carlo calculations for deuterium equations of state, and perform without
computational cost increase compared to zero-temperature functionals and so
should be used for finite-temperature calculations". 

Karasiev {\it et al.}~\cite{KaraQE}  have recently implemented the PDW-finite-$T$ XC functional
as well as  their new fit to the PIMC data in the `Quantum Espresso' code. They have
made calculations of the bandstructure and electrical conductivity of WDM Aluminum.
They find that the use of finite-$T$ XC is necessary if significant errors (upto 15\%
at $T/Ef\simeq$ 0.11 in the case of Al) are to  be avoided~\cite{KaraWDM}.

\subsection{Can we define free and bound electrons in an `unambiguous' manner?}
\label{free-bound.sub}
In a `fully-ionized' plasma all the electrons are in delocalized states. Thus, in
stark contrast to quantum chemistry, most of plasma physics deals with  continuum
processes. WDM systems usually contain some partially occupied bound states as
well as continuum states. Thus, if the Hamiltonian is bounded, and if there is no
frequency dependent external field acting on the system, there is no difficulty
in identifying the bound states and continuum states of the non-interacting
electron system used in Kohn-Sham theory. If a strong frequency-dependent
external field is acting on the system, the concept of `bound' electrons as
distinct from `free' electrons becomes much more hazy, and will not be discussed
here.

Depending on the nature of the `external potential', a system at $T=0$ may be
such that all electrons are in `bound states'. The latter are usually eigenstates
$\psi_j$  whose square $\psi_j(r)$  become rapidly negligible as $r$ goes beyond
a region of localization. The spectrum contains occupied and unoccupied  `bound
states' as  well as positive-energy states which are not localized within a given
region. All states become partially occupied in finite-$T$ systems, and
treatments that restrict themselves to a small basis set of functions localized
over a finite region of space become too restrictive. Most DFT codes use a
simulation cell of linear dimension $L$ with periodic boundary conditions. In
such a model the smallest value of $k$ in momentum space is $\sim \pi/L$ and
this prevents the direct evaluation of various properties (e.g., $S(k)$) as $k\to
0$. In the NPA model a  large sphere of radius $R$ such that all particle
correlations have died out is used, and phase shifts of continuum states, taken
as plane waves, are calculated. This procedure allows an essentially direct
access to $k\to 0$ properties as well as the bound and continuum spectrum of the
ion in the plasma. However, the difficulty arises when the electronic
bound-states spread beyond the Wigner-Seitz radius of the ion.

The question of determining the number of free electrons per ion, viz., $\bar{Z}$
is usually posed in the context of the mean-ionic charge $\bar{Z}$ used in metal
physics and  plasma theory. If the nuclear charge is $Z_n$, and if the total
number of bound electrons attributed to that nucleus is $n_b$, then clearly
$\bar{Z}=Z_n-n_b$ if the charge distribution $n_b(r)$ is fully contained
within the Wigner-Seitz sphere of the ion. 
While $n_b$ is well-defined in that sense for many elements under standard
conditions, giving, for example $\bar{Z}=3$ for Al at normal compression and up
to about $T$=20 eV,  this simple picture breaks down for many elements even under
normal conditions. If the electronic charge density cannot be accurately
represented as a superposition of individual atomic charge densities, the definition
of $n_b$ becomes more complicated since a bound electron may be shared between
two or more neighbouring atoms that form  bonds. Transition-metal solids and WDMs
have $d$-electron states which extend outside the atomic Wigner-Seitz sphere.
Hence assigning them to a particular nuclear center becomes a delicate exercise.
However, even in such situations there are meaningful ways to define $n_b$ and
$\bar{Z}$ that lead to consistence with experiment. In such situations the proper
value of $\bar{Z}$ may differ from one physical property to another as the
averaging  involved in constructing the mean value $\bar{Z}$ may change. A
similar situation applies to the effective  electron mass $m^*_e$ which deviates
from the ideal value of unity (in atomic units), and takes on different values
according to whether we are discussing a thermal mass, an optical effective mass,
or a band mass that we may use in a Luttinger-Kohn $k\cdot p$ calculation.

Experimentally, $\bar{Z}$ is a measure of the number of free electrons released
per atom. This can be measured from the $\omega\to 0$ limit of the optical
conductivity $\sigma(\omega)$. Thus, although transition metals like gold have
delocalized $d$-electrons, the static conductivity upto about 2 eV is found to
indicate that $\bar{Z}=1$, with the optical mass $m^*_e=1$. Another property
which measures $\bar{Z}$ is the electronic specific heat. Here again the specific
heat evaluated from DFT calculations that use a $\bar{Z}=1$ pseudopotential for
Au agrees with  experimental data up to 2 eV, while those that  use the density
of states from {\it  all 11 electrons} as free-electron  states will obtain
significantly different answers~\cite{Vermont,chen2013} that need to be used
with  circumspection. That is, such a calculation will be valid only if the
$d$-electrons are fully delocalized and partake in the heating process by being
coupled with the pump laser creating the WDM.

The argument that $\bar{Z}$ is not a valid concept or a quantum property
 because there is no `operator' corresponding to it has no merit.
 The temperature also does not correspond to the
mean value of a quantum operator. In fact, $T$ is a Lagrange multiplier ensuring the
constancy of the Hamiltonian within the relevant times scales, while $\bar{Z}$ is
the Lagrange multiplier that sets the charge neutrality condition 
$\bar{n}=\bar{Z}\bar{\rho}$ relating the average electron density to the
 average ion density~\cite{dwp82}.

Additional discussions regarding  $\bar{Z}$ may be found in
Refs.~\cite{cdw-CPP,ppots12} and in Ref.~\cite{pdw-Al} where the case of a WDM
mixture of ions with different ionization, viz., Al$^{Z_j+}$ is treated within a
first-principles DFT scheme.

\section{Future challenges in formulating finite-$T$ XC functionals}
\label{future.sec}
In considering a system of ions with a distribution
$\rho(\vec{r})=\sum_j\delta(\vec{r}-\vec{R}_j)$, and an electron distribution
interacting with it, the free energy $F$ has to be regarded as a functional of
both $\rho(r)$ and $n(r)$. Hence the ground state has to be determined by a
coupled variational problem involving a constrained-search minimization with
 respect to all physically possible electron charge distributions $n(r)$, and ion
 distributions $\rho(r)$, subject to
the usual formal constraints of $n$-representability etc. The Euler-Lagrange
variational equation from the derivative of $F$ with respect to $n(r)$, for a
fixed $\rho(r)$ would yield the usual Kohn-Sham procedure with the rigid
electrostatic potential of  $\rho(r)$ providing the external potential. However,
if no static approximation or Born-Oppenheimer approximation is made, we can
obtain another Euler-Lagrange  variational equation from the derivative of $F$
with respect to $\rho$. This coupled pair of equations treated via
density-functional theory involves not only the $f_{xc}^{ee}$, but also
$f_{xc}^{ei}$ and $f_c^{ii}$, the latter involving correlations (but no exchange)
as it arises from ion-ion interactions beyond the self-consistent-field
approximation. In effect, just as the electron many-body problem can be reduced
to an effective one-body problem in the  Kohn-Sham sense, we can thus reduced the
many-ion problem into a ``single-ion problem". Such an analysis was given by us
long ago~\cite{dwp82}.  

The ion-ion correlations cannot be
approximated by any type of local-density approximation, or even with a
sophisticated gradient approximation. However, Perrot and the present author
 were able to show that a fully
non-local approximation where an ion-ion pair-distribution can be constructed
{\it in situ} using the HNC  equation provides a very satisfactory solution.
This is equivalent to positing that the ion-ion correlation functional is made up
of the hyper-netted-chain diagrams. However, significant insights are
needed in regard to the electron-ion correlation functionals which involve
 the coupling
between a quantum subsystem and a classical subsystem. This is largely an open
problem that we have attempted to deal with via the classical-map HNC approach,
to be discussed below.

The advent of WDM and ultra-fast matter has thrown out a number of new challenges
to the implementation of thermal DFT. A simple but at the moment unsolved problem
in UFM may be briefly described as follows. A metallic solid like Al  at room
temperature ($T_r$) is subject to a short-pulse laser which heats the conduction
electrons to a temperature $T_e$ which may be 6 eV. The core electrons (which
occupy energy bands deep down in energy and hence not excitable by the laser)
remain essentially unperturbed in the core region and at the core temperature,
i.e., at $T_r\simeq$ 0.026 eV. The temperature  relaxation by electron-ion
processes is `slow', i.e., it occurs in pico-second times scales. On  the other
hand, electron-electron processes are `fast', and hence one would expect that the
conduction-band electrons  at $T_e$ to undergo exchange as well as Coulomb
scattering within femto-second time scales, consistent  with electron-electron
interactions timescales. Thus, while we have a quasi-equilibrium of  a
two-temperature system holding for up to pico-second timescales, the question
arises if one can meaningfully calculate an exchange and correlation potential
between the bound electrons in the core at the temperature $T_r$, and the
conduction-band electrons at $T_e$, with $T_e\gg T_i$. While we believe on
physical grounds that a thermal DFT is applicable at least in an approximate
sense, an unambiguous method for calculating the two-temperature XC-energies and
potentials is  as yet unavailable. 
\subsection{Classical-map Hyper-Netted Chain Method}
\label{chnc.sub}
Once the pair-distribution function of a classical or quantum Coulomb system is
known, all the thermodynamic functions of the system  can be calculated from
$g(r)$. The XC-information is also in the $g(r)$.  Only the ground-state
correlations are needed in calculating the linear transport properties of the
system. Hence, most properties of the system become available. It is well known
that correlations among classical charges (i.e, ions) can be treated with good
accuracy via the the hyper-netted-chain equation, but dealing with the quantum
equivalent of 
hyper-netted-chain diagrams for quantum systems is difficult, even at $T=0$
~\cite{FermiHNC}.

When we have an electron subsystem interacting with the ion subsystem, obtaining
the PDFs becomes a difficult quantum problem even via more standard
 methods. We need to solve for a
many-particle wavefunction  which rapidly becomes intractable as the number of
electrons is increased beyond a small number. The message of DFT is that the
many-body wavefunction is not needed, and that the one-particle charge
distribution $n(r)$ is sufficient. While the charge distribution at $T=0$
involves a sum over the squares of the occupied Kohn-Sham wavefunctions, at very
high $T$ the classical charge distribution is given by a Boltzmann distribution
containing an effective potential felt by a single `field'  particle, and
characterized by the temperature which is directly proportional to the classical
kinetic energy.

In CHNC we attempt to replace the
quantum-electron problem by a classical Coulomb problem where we can use a simple
method like the ordinary HNC equation to directly obtain the needed PDFs, at some
effective `classical fluid' temperature $T_{cf}$ having the same density
distribution as the quantum fluid. The electron PDF $g^0(r)$ of the
non-interacting quantum electron fluid is known at any temperature and embodies
the effect of quantum statistics (Pauli principle). Hence we can ask for the
effective potential $\beta V_{Pau}(r)$ which, when used in the HNC, gives us the
$g^0(r)$, an idea dating back to a publication by F. Lado~\cite{Lado}. This ensures
 that the non-interacting  density has the required 
``$n$-representable" form of a Slater determinant. Of course, only the
 product $P(r)=\beta V_{Pau}(r)$
can be determined by this method, and it exists even at $T=0$. Then the total
pair potential to be used in the equivalent classical fluid 
is taken as $\beta \phi(r)=P(r)+\beta V_{Cou}(r)$. How does
 one choose $\beta=1/T_{cf}$
since the Pauli term is independent of it?

\begin{figure}[t]
\centering
\includegraphics[width=10cm]{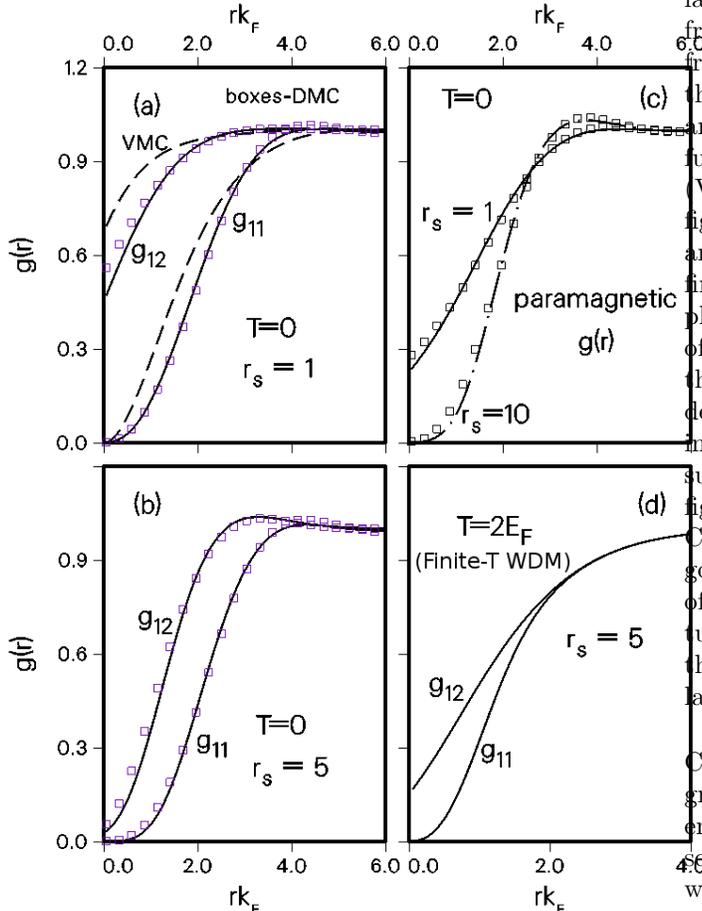}
\caption
{(a) Here the  CHNC g(r) are compared with VMC and DMC simulation results:
the interacting PDFs $g_{11}(r)$ and
$g_{12}(r)$ at $r_s$=1 are shown. Solid lines:-CHNC, boxes:-DMC,
dashed line:-VMC~\cite{vms-dms}.  Panel (b) $r_s=5$, DMC~\cite{vms-dms}
and HNC. In (c) the paramagnetic
$g(r)$ at $r_s$=1 and $r_s$=10, T=0 are compared with DMC.
(d)Finite temperature PDFs (CHNC) for $T/E_F$=2, $r_s$=5 would
correspond to a WDM at $\simeq$ 3.6 eV ($\sim$ 42,000 K).
}
\label{gr-fig}
\end{figure}

To a very good approximation, if  $T_{cf}$ is chosen such that the classical 
fluid has the same Coulomb correlation energy $E_c$ as the quantum electron fluid, then
it is found that the PDF  of the classical Coulomb fluid is a very close
approximation to the PDF of the quantum electron fluid at $T=0$. There is of
course no mathematical proof of this. However, from DFT we know that
only the `correct' ground state distribution will give us the correct energy, and
perhaps it is not surprising that this choice is found to work. The $T_{cf}$
that works for the $T=0$ quantum electron gas is called  the ``quantum temperature"
$T_q$. More details of the method are given in Ref.~\cite{pdw2000}. There it is
argued that, to a good approximation, for a  finite-$T$ electron gas at the
physical temperature $T$, the effective classical fluid temperature
$T_{cf}=\sqrt{T_q^2+T^2}$. This has been confirmed independently
by Datta and Dufty~\cite{SPanDufty} in their study of classical approximations
to the quantum electron fluid. Thus CHNC provides  all the tools necessary for
implementing a classical HNC calculation of the PDFs of the quantum electron gas
at finite-$T$. 

We display in Fig.~\ref{gr-fig} pair-distribution
functions calculated using CHNC, and those available in the literature from
quantum simulations at $T=0$, as finite-$T$ PDFs from quantum
simulations are hard to find. In any case the classical map is expected to be
better as $T$ increases and the $T=0$ comparison is important.
 In the figure, diffusion Monte Carlo (DMC) and variational
Monte Carlo (VMC) data\cite{vms-dms}  are compared with CHNC results.
In  fig.~\ref{gr-fig} the parallel-spin PDF is marked $g_{11}(r)$,
 while the anti-parallel
spin PDF is marked $g_{12}(r)$. The latter has a finite value as $r\to 0$
as there is no Pauli exclusion principle operating on them. Furthermore, the
the mean value of the operator of the Coulomb potential, i.e.,  $e^2/r$,
 is of the form
$\{1-\exp(-k_{dB} r)\}/r$, where $k_{dB}$ is the thermal de Broglie wavelength of
the electron pair, as discussed in Ref.~\cite{prl1}. This `quantum-diffraction'
correction
ensures that $g_{12}(r\to 0)$ has a finite value, as seen in the figure. It is in
good agreement with Quantum Monte Carlo results. Thus the CHNC is capable of
providing a good interpretation of the physics underlying the results of
quantum simulations.   Needless to say, unlike Quantum Monte Carlo or Path-Integral
simulation methods, the CHNC integral equations can be implemented on a
laptop and the computational times are imperceptible.

Using the PDFs $g(r,T,\lambda)$ calculated with a scaled
Coulomb potential $\lambda V_{Cou}(r)$, a coupling constant integration over
$\lambda$ can be carried out to obtain the XC-free energy $F_{xc}(r_s,T)$ as
described in detail in Ref.~\cite{pdw2000}. As seen from Fig.~\ref{fxc-fig},
this  procedure leads to good agreement with the thermal-XC results from the PIMC
method, while only the $T=0$ spin-polarized $E_c$ data were used in constructing $T_q$.
Furthermore, since $T_{cf}$ tends to the physical temperature at high $T$, and
since the HNC provides an excellent approximation to the PDFs of the high-$T$
electron system, the method naturally recovers the high-$T$ limit of the
classical one-component plasma. Note that we could NOT have started from the
high-$T$ limit of an ideal classical gas and used the well-known classical
coupling constant (i.e., $\Gamma$ integration method, e.g., see Baus
and Hansen or Ichimaru~\cite{BausHan80, Ichimaru}) 
to determine $f_{xc}$ from an integration that
ranges from $\Gamma=0, T=\infty$ to the needed temperature  (i.e, the needed
$\Gamma$). This is because  there is no clear method of capturing the physics
contained in $T_q$,  and ensuring that Fermi statistics are obeyed
(e.g., via the introduction of a
$\beta V_{Pau}(r)$), as there is only Boltzmann statistics at $\Gamma=0$. 

The ability of the CHNC to correctly capture the thermal-DFT properties of the
finite-$T$ quantum fluid suggests its use for electron-ion systems like
compressed hydrogen (electron-proton gas), or complex plasmas with many different
classical ions interacting with electrons~\cite{hug}, {\it without having to
solve the Kohn-Sham equations} as in Bredow {\it et al.}~\cite{Bredow15}. The
 extensive calculations of Bredow {\it et al.} establish the ease and rapidity 
provided by CHNC, without sacrificing accuracy.
CHNC has potential applications for
electron-positron systems or electron-hole systems where both quantum components
can be treated via the classical map. It also provides a partial solution to the
still unresolved problem of formulating a fully-nonlocal `orbital-free' approach
 that directly
exploits the Hohenberg-Kohn-Mermin theory, without the need to go via the
Kohn-Sham orbital formulation. 

\section{Conclusion}
We have argued that our current knowledge of the thermal XC-functionals is
satisfactory and the stage is set for their  implementation in practical DFT
codes. Noting the complexity of warm-dense matter, we have emphasized
simplifications as well as extensions which do not sacrifice accuracy. In this
respect the neutral-pseudo atom model can, in most cases do the work of the
{\it ab initio} codes like VASP, and handle high-temperature problems that are
beyond their scope. Orbital-free approaches~\cite{KaraQE} will also become
 increasingly useful, especially at intermediate and high $T/E_F$.
Nevertheless,  the {\it ab initio} codes are needed at
low-temperature low-density situations involving molecular formation,
where the NPA breaks down as it is a
``single-center'' approach. However, in many WDM cases, we need to go beyond
the picture where the ion subsystem is held static, and the electrons only
feel them as an `external potential'. Hence we have emphasized the need for
calculating not just the XC-functionals for electrons, but also the classical
correlation functionals for ions, as well as the ion-electron correlations
directly,  {\it in situ}, via direct coupling-constant integrations of all the
pair-distribution functions of the system, ensuring a fully non-local formulation
where gradient expansions are not needed. In fact, there is no need for any
XC-functionals in such a scheme. To do this efficiently and 
accurately, we have proposed a classical map of the quantum electrons and
implemented it in the CHNC scheme which depends on DFT ideas. This capacity is not
found in any of the currently available methods. CHNC has been
 used to construct a finite-$T$
XC functional for electrons more than a decade before PIMC results became
available, and it turns out that the CHNC results are accurate. 
The CHNC scheme has been successfully used for calculating  the
equation of state and other properties of warm dense matter as well as
multi-component $T=0$ electron-layer systems, thick layers etc.,
 that are expensive to treat by quantum
 simulation methods, but relevant for nanostructure physics.

\end{document}